\documentclass[aps,prx,twocolumn,floatfix,longbibliography,nofootinbib]{revtex4-2}
\usepackage[utf8]{inputenc}
\usepackage{natbib}
\usepackage{graphicx}
\usepackage{xcolor}
\usepackage[colorlinks=true, linkcolor=blue]{hyperref}
\usepackage{amssymb}
\usepackage{gensymb}
\usepackage{comment}

\usepackage{newpxmath}
\usepackage{float}
\usepackage{amsmath}
\usepackage{tabularx}
\usepackage{epstopdf}
\usepackage{latexsym}
\usepackage{color, colortbl}
\usepackage{psfrag}
\usepackage{bbm}
\usepackage{bm}
\usepackage{titlesec}
\usepackage{dsfont}
\usepackage{feynmp}
\usepackage{slashed}
\usepackage{subcaption}
\usepackage{multirow}
\usepackage[newcommands]{ragged2e}
\usepackage[normalem]{ulem}

\usepackage{braket}
\usepackage{relsize}

\def \beq {\begin{eqnarray}}
\def \eeq {\end{eqnarray}}

\usepackage{amsfonts,bm}

\newcommand{\kb}{k_{\mathrm B}}
\begin{document}
\title{Information, Dissipation, and Planckian Optimality}
\author{Debanjan Chowdhury}\email{debanjanchowdhury@cornell.edu}
\affiliation{Department of Physics, Cornell University, Ithaca NY 14853}
\date{\today}
\begin{abstract}
We derive a universal bound on the efficiency with which ``dissipated" work can generate distinguishable changes in a quantum many-body state at a finite temperature, as quantified by the quantum Fisher information. The bound follows solely from the analytic structure of equilibrium many-body correlators and is independent of all microscopic details. It takes a frequency-resolved form with a characteristic crossover at the Planckian scale, $\omega_\star\sim \kb T/\hbar$. We find that Planckian scatterers sit at the edge of optimality, displaying maximal relaxation rate before information-dissipation efficiency collapses. This suggests strange metals are not just fast dissipators, but the fastest that remain efficient in generating distinguishability. The bounded quantity can be evaluated directly from optical conductivity measurements in strongly correlated electronic systems, offering a unique window into how dissipation generates distinguishable changes.
\end{abstract}
\maketitle

\textit{Introduction}--- A central problem in quantum many-body physics is understanding the subtle interplay of information and dissipation in strongly interacting, time-dependent systems.
In closed quantum systems, time-dependent drives generically produce dissipation at the level of coarse-grained observables, redistributing energy among many-body degrees of freedom~\cite{DAlessio2016,Mori2018}. Although we only study unitary dynamics, we will use ``dissipation'' to denote energy transferred from the controlled drive into complex many-body degrees of freedom. The energy transferred cannot be recovered simply by reversing the control protocol and thus represents an operationally irreversible cost~\cite{Jarzynski2011,Esposito2010}.
This raises a fundamental question: \emph{how efficiently can dissipation generate distinguishable changes in a quantum many-body state at a finite temperature?}

In this Letter, we establish a universal bound connecting linear-response theory of thermal many-body systems to the geometry of quantum states.
Our central quantity is the \emph{Bures distance}~\cite{Bures1969, Uhlmann1976}, related to the \emph{quantum Fisher information}  (QFI)~\cite{Helstrom1976,Braunstein1994}, which quantifies the optimal distinguishability of nearby quantum states. 
Applied to driven many-body systems, the QFI provides an operational measure of how effectively a perturbation is imprinted onto a thermal state.
We show that the Bures distance (and relatedly, QFI) generated by a drive is universally bounded by the entropy production during the same process.
This bound follows solely from Kubo-Martin-Schwinger (KMS) relations~\cite{Kubo1957,Martin1959} and positivity conditions, independent of microscopic Hamiltonian details, and requires no assumptions about weak coupling, quasiparticles, or specific scattering mechanisms. Remarkably, the bound takes a frequency-resolved form with a characteristic crossover at the Planckian scale. 

\begin{figure}
\centering
\includegraphics[width= \columnwidth]{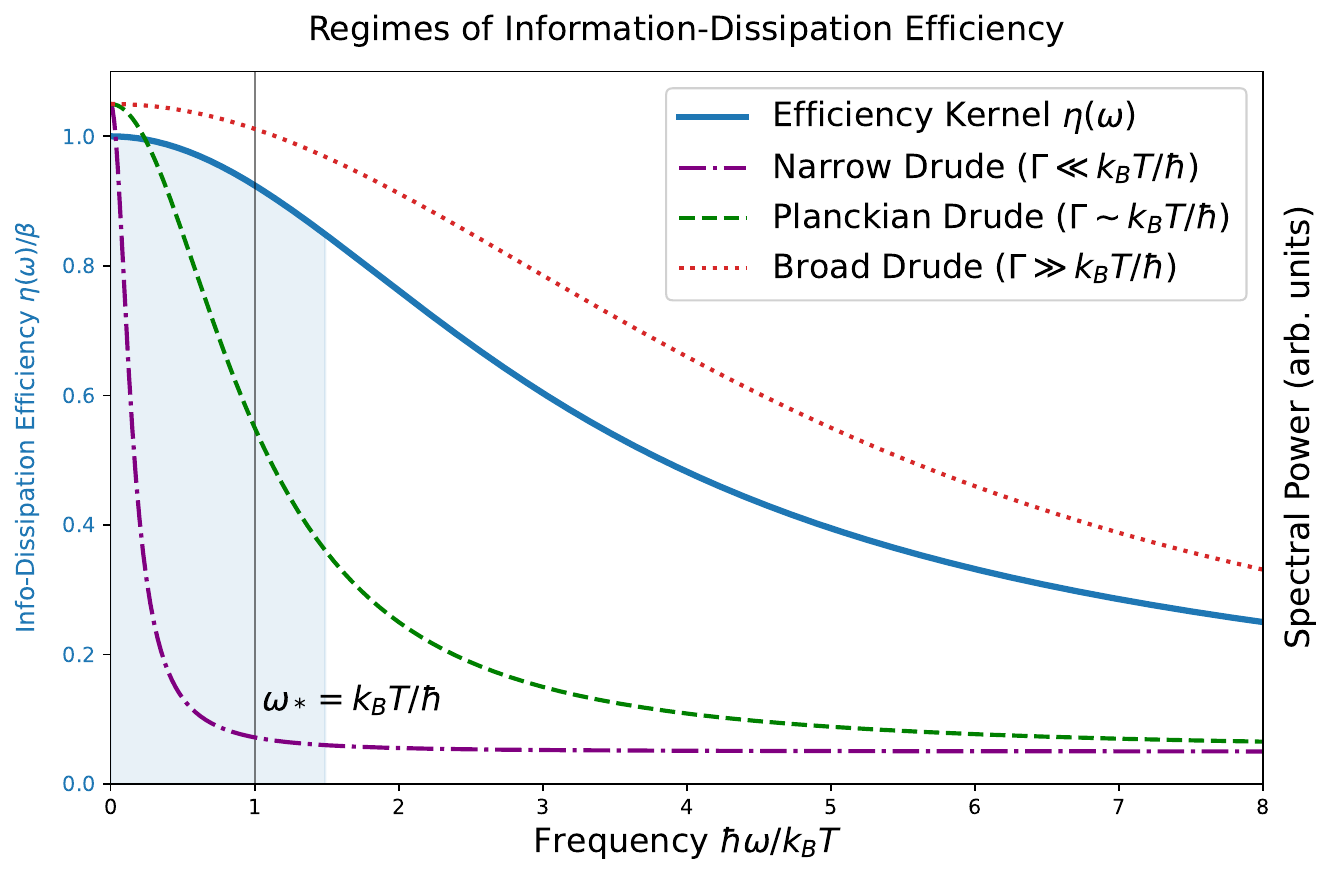}
\caption{Normalized information-dissipation efficiency kernel (solid blue curve), $\eta(\omega)/\beta$; see Eq.~\eqref{eq:main_bound_intro}. The shaded region indicates the approximate Planckian window where dissipation is maximally efficient. Three representative dissipative optical spectra are shown: a narrow Drude 
(purple dash-dotted), a Planckian Drude (green dashed), and a broad Drude (red dotted) peak, respectively. Systems with spectral weight concentrated in the Planckian window (shaded region) achieve near-optimal information-dissipation efficiency, while those 
with significant high-frequency weight are parametrically inefficient.}
\label{main_fig}
\end{figure}
Defining the squared Bures distance, $d_B^2$, between the driven and equilibrium states, we establish
\begin{equation}
\frac{d_B^2}{W_{\mathrm{diss}}} = \frac{\langle \eta(\omega) \rangle_{P(\omega)}}{4}\leq \frac{\beta}{4},
\label{eq:main_bound_intro}
\end{equation}
which directly bounds state distinguishability by dissipated work ($W_{\mathrm{diss}}$), with no extraneous parameters. The efficiency of information encoding is governed by a universal kernel $\eta(\omega) = (2/\hbar\omega)\tanh(\beta\hbar\omega/2)$, which interpolates between $\beta$ at low frequencies and $2/\hbar\omega$ at high frequencies, with a crossover at the Planckian scale, $\omega_\star \sim k_B T/\hbar$; see Fig.~\ref{main_fig}.  The expectation value is computed w.r.t. a normalized probability distribution, $P(\omega)$, governed by the drive protocol, as we discuss below. This analytical structure reveals that dissipation occurring below the Planckian frequency is maximally efficient at generating distinguishability, while high-frequency dissipation is parametrically suppressed.

Our results connect to several lines of inquiry in a complementary fashion. Thermodynamic uncertainty 
relations~\cite{Barato2015,Gingrich2016,Hasegawa2020} establish precision-entropy 
trade-offs but apply to steady-state currents, not transient response. 
Quantum speed limits~\cite{Mandelstam1945,Margolus1998,Deffner2017} bound 
evolution rates via energy uncertainty; our bound involves dissipation and 
thermal distinguishability with no energy-variance dependence. Prior information-geometric bounds~\cite{Deffner2010,Plastina2014} constrain 
entropy production given a target state; we bound QFI along a control 
direction given fixed dissipation. Equilibrium uncertainty 
relations~\cite{MengShi2025} connect QFI to static susceptibilities for 
conjugate variables; our bound involves dynamical driving and dissipated work. A unifying distinction is that our bound is frequency-resolved and the KMS 
structure introduces a universal kernel $\eta(\omega)$ exposing the Planckian 
scale as a crossover between efficient and inefficient dissipation. 
Near-saturation reflects spectral concentration below the Planckian frequency, not proximity to quasistatic or equilibrium limits.

One motivation for this work stems partly from the empirical observation of Planckian phenomenology \cite{JZaanen2004,SSachdev1999b} in correlated quantum materials~\cite{JANBruin2013,legrosUniversalTlinearResistivity2019a,YCao2020,AJaoui2022,Paglione,AP21,Grissonnanche2020}, where transport relaxation times are governed by $O(\omega_\star^{-1})$.
Such observations are typically phrased as bounds on scattering rates, but these interpretations are probe-dependent and rely on model-specific fitting procedures \cite{SAHartnoll2022,DChowdhury2022a}. Theoretical progress has been made by proposing bounds on the charge diffusivity \cite{SAHartnoll2015,MBlake2016,ALucas2016,lucashartnoll17,hartnoll17,hartnoll18,Luca25}, Lanczos coefficients \cite{auerbach18,DEParker2019}, uncertainty relations \cite{nussinov}, and spectral moments of the optical conductivity \cite{DC25}. 
Our bound in the present manuscript offers a complementary perspective: the Planckian timescale emerges as the frequency below which dissipation most efficiently encodes information (i.e. distinguishability) into a thermal state.
Near-saturation of the bound, when dissipative spectral weight is concentrated below $\omega \lesssim k_{\mathrm{B}}T/\hbar$, provides an operational, model-independent characterization of Planckian ``optimality'' that disentangles kinematic constraints imposed by thermal quantum geometry from dynamical scattering mechanisms; see Fig.~\ref{main_fig}.

\textit{Distinguishability \& QFI}---
Consider a closed quantum many-body system with Hamiltonian $H_0$ in thermal equilibrium at inverse temperature $\beta = (k_{\mathrm{B}} T)^{-1}$, described by the Gibbs state $\rho_\beta = e^{-\beta H_0}/Z$ with partition function $Z = \mathrm{Tr}(e^{-\beta H_0})$.
We couple a spatially uniform, time-dependent vector potential $A(t)$ to the total current operator $J$ via minimal coupling,
\begin{equation}
H(t) = H_0 - A(t)\, J,
\label{eq:H_driven}
\end{equation}
and parametrize $A(t) = \theta\, \lambda(t)$, where $\lambda(t)$ is a dimensionless waveform that vanishes at late times, and $\theta$ carries units of vector potential.
The goal is to quantify how much information about the parameter $\theta$ is encoded into the evolved quantum state.

The natural measure of parameter distinguishability is the QFI~\cite{Helstrom1976,Braunstein1994}.
For a family of states $\{\rho_\theta\}$, the QFI $F_Q(\rho_\theta)$ determines the ultimate precision with which $\theta$ can be estimated via the quantum Cram\'{e}r-Rao bound, $\mathrm{Var}({\theta}) \geq 1/F_Q$~\cite{Holevo1982}.
For mixed states, several inequivalent QFI metrics exist; we employ the Bures (or symmetric logarithmic derivative) metric~\cite{Bures1969,Uhlmann1976}. The Bures QFI is related to the Bures distance $d_B$ introduced previously, a Riemannian metric on the space of density matrices, via $d_B^2(\rho_\theta, \rho_0) = \frac{1}{4} F_Q \theta^2 + O(\theta^3)$ for small $\theta$~\cite{Braunstein1994, Toth2014}. The Bures distance provides a basis-independent measure of state distinguishability, with $d_B = 0$ if and only if the states are identical. The QFI has found wide application in quantum metrology~\cite{Giovannetti2006, Pezze2018} and, more recently, as a probe of many-body entanglement accessible through dynamic susceptibilities~\cite{Hauke2016,konik24,paschen24,abbamonte25}.

Working in the interaction picture with respect to $H_0$, the current operator evolves as $J(s) = e^{i H_0 s/\hbar} J\, e^{-i H_0 s/\hbar}$, and to leading order in $\theta$ the time-evolution operator takes the form
\begin{equation}
U_\theta(t) = e^{-i H_0 t/\hbar} \left[ \mathbb{I} + \frac{i\theta}{\hbar} K_t + O(\theta^2) \right],
\end{equation}
where the integrated generator is
\begin{equation}
K_t = \int_0^t ds\, \lambda(s)\, J(s).
\label{eq:generator}
\end{equation}
Let $\{\ket{n}\}$ denote the eigenbasis of $H_0$ with energies $E_n$ and Boltzmann weights $p_n = e^{-\beta E_n}/Z$.
The QFI at $\theta = 0$, denoted $F_Q(t) := F_Q(\rho_\theta(t))|_{\theta=0}$, is given by~\cite{Paris2009,QFI12a,QFI12b,QFI14},
\begin{equation}
F_Q(t) = \frac{2}{\hbar^2} \sum_{m,n} \frac{(p_m - p_n)^2}{p_m + p_n} \left| \bra{m} K_t \ket{n} \right|^2.
\label{eq:FQ_spectral}
\end{equation}
This expression makes manifest that only off-diagonal matrix elements connecting states with different Boltzmann weights contribute to distinguishability.

To obtain a frequency-domain representation, we use the identity
\begin{equation}
\frac{(p_m - p_n)^2}{p_m + p_n} = (p_m + p_n) \tanh^2\!\left( \frac{\beta(E_m - E_n)}{2} \right),
\label{eq:tanh_identity}
\end{equation}
which yields
\begin{subequations}
\label{eq:FQ_frequency}
\begin{align}
F_Q(t) &= \int_{-\infty}^{\infty} \frac{d\omega}{2\pi}\, |\tilde{\lambda}(\omega)|^2\, \mathcal{K}_B(\omega), \\
\mathcal{K}_B(\omega) &= \frac{2}{\hbar^2} \tanh^2\!\left( \frac{\beta\hbar\omega}{2} \right) S_{JJ}(\omega),
\end{align}
\end{subequations}
where $\tilde{\lambda}(\omega) = \int dt\, e^{i\omega t} \lambda(t)$ is the Fourier transform of the drive waveform.
Here, $S_{JJ}(\omega)$ is the symmetrized current spectral function,
\begin{equation}
S_{JJ}(\omega) = \int dt\, e^{i\omega t}\, \frac{1}{2} \mathrm{Tr}\!\left( \rho_\beta \{J(t), J(0)\} \right) \geq 0,
\label{eq:symmetrized_spectrum}
\end{equation}
which is positive semi-definite by construction.
The kernel $\mathcal{K}_B(\omega)$ encodes how thermal fluctuations at frequency $\omega$ contribute to distinguishability: the $\tanh^2(...)$ factor suppresses contributions at low frequencies $\omega \ll \omega_\star$, reflecting the fact that thermal states with nearby energies are intrinsically difficult to distinguish.

\textit{Dissipation in linear response}---
Although time evolution under Eq.~\eqref{eq:H_driven} is unitary, the work performed by the external drive has an irreversible component that constitutes dissipation in the standard thermodynamic sense~\cite{Jarzynski1997,Crooks1999}.
Within linear response, this dissipated work $W_{\mathrm{diss}}(t)$ is governed by the absorptive part of the response function.

The instantaneous power delivered by the drive is
\begin{equation}
P(t) = \mathrm{Tr}\!\left( \rho_\theta(t)\, \partial_t H(t) \right) = -\theta\, \dot{\lambda}(t)\, \langle J \rangle_{\theta,t}.
\end{equation}
To leading order in $\theta$, the current expectation value follows from linear response~\cite{Kubo1957},
\begin{equation}
\langle J \rangle_{\theta,t} = \theta \int_0^t ds\, \chi^R_{JJ}(t-s)\, \lambda(s) + O(\theta^2),
\label{eq:current_response}
\end{equation}
where the retarded current-current susceptibility is
\begin{equation}
\chi^R_{JJ}(t) = \frac{i}{\hbar} \Theta(t)\, \mathrm{Tr}\!\left( \rho_\beta [J(t), J(0)] \right).
\label{eq:chi_retarded}
\end{equation}
Transforming to the frequency domain, the dissipated work at $O(\theta^2)$ takes the form,
\begin{subequations}
\label{eq:Wdiss}
\begin{align}
W_{\mathrm{diss}}(t) &= \theta^2 \int_{-\infty}^{\infty} \frac{d\omega}{2\pi}\, |\tilde{\lambda}(\omega)|^2\, \mathcal{K}_{\mathrm{diss}}(\omega), \\
\mathcal{K}_{\mathrm{diss}}(\omega) &= \omega\, \chi''_{JJ}(\omega) \geq 0,
\end{align}
\end{subequations}
where $\chi''_{JJ}(\omega) = \mathrm{Im}\, \chi^R_{JJ}(\omega)$ is the dissipative (absorptive) part of the response, and positivity of $\mathcal{K}_{\mathrm{diss}}(\omega)$ for $\omega>0$ follows from Lehmann representation. The associated entropy production is $\Delta S_{\mathrm{irr}} = \beta W_{\mathrm{diss}}$ 
(in units where $k_{\mathrm{B}} = 1$).

The connection between fluctuations and dissipation is encoded in the fluctuation-dissipation theorem~\cite{Callen1951,Kubo1966},
\begin{equation}
S_{JJ}(\omega) = \hbar\, \coth\!\left( \frac{\beta\hbar\omega}{2} \right) \chi''_{JJ}(\omega),
\label{eq:FDT}
\end{equation}
which relates the symmetrized fluctuation spectrum $S_{JJ}(\omega)$ appearing in the QFI kernel [Eq.~\eqref{eq:FQ_frequency}] to the dissipative response $\chi''_{JJ}(\omega)$ controlling work absorption.
This relation, a direct consequence of the KMS condition for thermal equilibrium~\cite{Haag1967}, enables us to derive a universal bound connecting distinguishability and dissipation.

\textit{Statement of the bound}--- Combining Eqs.~~\eqref{eq:FQ_frequency}, \eqref{eq:Wdiss}, and \eqref{eq:FDT}, we can form the ratio of the QFI and dissipation kernels:
\begin{equation}
\eta(\omega) := \frac{\mathcal{K}_B(\omega)}{\mathcal{K}_{\mathrm{diss}}(\omega)} = \frac{2}{\hbar\omega} \tanh\frac{\beta\hbar\omega}{2}.
\label{eta}
\end{equation}
This \emph{information-dissipation efficiency} (Fig.~\ref{main_fig}) is a universal function depending only on the dimensionless ratio $\beta\hbar\omega$, with no reference to microscopic details of $H_0$.

Since $\tanh x \leq x$ for all $x \geq 0$, with equality only as $x \to 0$, we obtain $\eta(\omega) \leq \beta$ uniformly. The QFI can be expressed as
\begin{equation}
F_Q(t) = \left\langle \eta(\omega) \right\rangle_{P(\omega)} \cdot \frac{W_{\mathrm{diss}}(t)}{\theta^2},
\end{equation}
where
\begin{equation}
P(\omega) = \frac{|\tilde{\lambda}(\omega)|^2 \mathcal{K}_{\mathrm{diss}}(\omega)}{\int \frac{d\omega}{2\pi} |\tilde{\lambda}(\omega)|^2 \mathcal{K}_{\mathrm{diss}}(\omega)}
\end{equation}
is a normalized probability distribution representing the dissipation-weighted spectral density of the drive protocol. Using $d_B^2 = \frac{1}{4} F_Q \theta^2$, the bound $\eta(\omega) \leq \beta$ yields our central result in Eq.~\eqref{eq:main_bound_intro}, which can be expressed in terms of $F_Q$ as
\begin{equation}
\frac{F_Q}{W_{\mathrm{diss}}/\theta^2} = \langle \eta(\omega) \rangle_{P(\omega)} \leq \beta.
\label{eq:bound}
\end{equation}
Note that while $\theta$ appears in the parametrization, both $F_Q$ and 
$ W_{\mathrm{diss}}/\theta^2$ are independent of $\theta$; the former depends only on the drive waveform and equilibrium fluctuations, while the latter is the dissipation per unit drive intensity. The bound is thus an intrinsic property of the system and protocol.

The frequency structure of $\eta(\omega)$ reveals how the Planckian scale emerges:
\begin{equation}
\eta(\omega) = 
\begin{cases}
\beta \left[ 1 - \dfrac{(\beta\hbar\omega)^2}{12} + O((\beta\hbar\omega)^4) \right], & \beta\hbar\omega \ll 1, \\[8pt]
\dfrac{2}{\hbar\omega}, & \beta\hbar\omega \gg 1.
\end{cases}
\label{eq:eta_limits}
\end{equation}
At low frequencies $\omega \ll k_{\mathrm{B}}T/\hbar$, the efficiency saturates at its maximum value $\beta$, i.e. every unit of dissipated work is converted into distinguishability with optimal efficiency.
At high frequencies $\omega \gg \omega_\star$, the efficiency decays as $1/\omega$, and dissipation becomes parametrically inefficient at generating QFI.
The crossover occurs at the Planckian scale $\omega_{\star}$. The bound in Eq.~\eqref{eq:main_bound_intro} is saturated in the limit where the dissipation-weighted spectrum $P(\omega)$ is supported entirely at $\omega \to 0$.
Near-saturation occurs when $P(\omega)$ concentrates predominantly in the Planckian window $\omega \lesssim \omega_\star$.
This provides an operational characterization of Planckian behavior, namely that systems that dissipate primarily at frequencies below the thermal scale achieve near-optimal information-dissipation efficiency.

The bound in Eq.~\eqref{eq:main_bound_intro} thus states that the squared Bures distance, quantifying how distinguishable the driven state is from equilibrium, cannot exceed $\beta/4$ times the dissipated work. Equivalently, using $\Delta S_{\mathrm{irr}} = \beta W_{\mathrm{diss}}$,
\begin{equation}
d_B^2 \leq \frac{1}{4} \Delta S_{\mathrm{irr}},
\end{equation}
bounding distinguishability directly by entropy production.

Via the Cram\'er-Rao inequality $\mathrm{Var}({\theta}) \geq 1/F_Q$ and $F_Q = 4 d_B^2/\theta^2$, this implies a thermodynamic lower bound on estimation precision:
\begin{equation}
\mathrm{Var}({\theta}) \geq \frac{k_B T \, \theta^2}{W_{\mathrm{diss}}}.
\end{equation}
While in the many-body setting, one is not necessarily trying to estimate $\theta$, the lower bound  quantifies a fundamental cost of precision: to estimate the perturbation strength $\theta$ more accurately, one must either dissipate more work or operate at lower temperature. This trade-off is reminiscent of thermodynamic uncertainty relations~\cite{Barato2015,Horowitz2020}, and relies on the KMS conditions involving the QFI rather than current fluctuations.

Several clarifications are now in order.
First, the bound constrains information-dissipation efficiency, \emph{not} microscopic scattering rates; it is compatible with any form of $\chi''_{JJ}(\omega)$.
Second, unlike Mandelstam-Tamm-type quantum speed limits~\cite{Mandelstam1945,Deffner2017}, which bound evolution rates in terms of energy uncertainty $\Delta E$, our bound is controlled by irreversibility and temperature, with no explicit dependence on the energy scale of $H_0$.
Third, the bound is intrinsically thermal and is constrained by the KMS structure of equilibrium states.

\textit{Electric field \& optical conductivity}---
The bound in Eq.~\eqref{eq:main_bound_intro} can be recast in terms of directly measurable optical response functions, facilitating connection to experiment. Defining the electric field $E(t) = -\dot{A}(t)$, the dissipative part of the optical conductivity is related to the current susceptibility by~\cite{Mahan2000, Dressel2002},
\begin{equation}
\mathrm{Re}\, \sigma(\omega) = \frac{1}{\omega} \chi''_{JJ}(\omega) \geq 0.
\label{eq:optical_conductivity}
\end{equation}
Introducing a normalized waveform $e(t)$ via $E(t) = \theta \, e(t)$, where $\tilde{e}(\omega) = i\omega \tilde{\lambda}(\omega)$, the Bures distance and dissipated work become
\begin{align}
d_B^2 &= \frac{1}{4} \int_0^\infty \frac{d\omega}{2\pi} |\tilde{e}(\omega)|^2 \, \mathrm{Re}\,\sigma(\omega) \, \eta(\omega), \\
W_{\mathrm{diss}} &= \int_0^\infty \frac{d\omega}{2\pi} |\tilde{e}(\omega)|^2 \, \mathrm{Re}\,\sigma(\omega),
\end{align}
where the second line is the standard Joule heating formula~\cite{Landau1984}. Both quantities depend only on the waveform shape $e(t)$ and equilibrium response. The bound can be reinterpreted as constraining the $\eta$-average weighted by the \emph{absorbed spectral power} $|\tilde{e}(\omega)|^2 \mathrm{Re}\,\sigma(\omega)$,
\begin{equation}
\left\langle \eta(\omega) \right\rangle_\sigma \leq \beta.
\end{equation}

This formulation connects directly to optical spectroscopy.
Given a measured conductivity spectrum $\mathrm{Re}\, \sigma(\omega)$ and a choice of probe pulse $E(t)$, one can compute both quantities appearing in the bounded ratio and evaluate the efficiency.
Systems exhibiting Planckian transport are characterized by optical conductivity with spectral weight concentrated at low frequencies $\omega \lesssim k_{\mathrm{B}}T/\hbar$, often described by a Drude form with width $\Gamma \sim k_{\mathrm{B}}T/\hbar$~\cite{orenstein,Marel2003,basov07,YRS20,Michon2023,DChaudhuri2025,LiangWu}.
For such systems, the absorbed power spectrum naturally concentrates in the Planckian window, leading to near-saturation of the bound---an operational signature of Planckian optimality distinct from (though related to) statements about scattering rates. Conversely, systems with significant spectral weight at high frequencies, $\omega \gg k_{\mathrm{B}}T/\hbar$, will exhibit sub-optimal information-dissipation efficiency.
The ratio $\langle \eta(\omega) \rangle_\sigma / \beta$ thus provides a dimensionless measure of how ``Planckian'' a material's dissipative response is, accessible entirely through optical measurements.

To make the bound concrete, consider a system with Drude-form conductivity,
\begin{equation}
\mathrm{Re}\, \sigma(\omega) = \frac{\sigma_0}{1 + (\omega/\Gamma)^2},
\label{eq:drude}
\end{equation}
where $\Gamma$ is the relaxation rate.
For monochromatic driving at frequency $\omega_0$, the efficiency is simply $\eta(\omega_0)/\beta$, which equals unity for $\omega_0 \ll k_{\mathrm{B}}T/\hbar$ and decays as $2k_{\mathrm{B}}T/\hbar\omega_0$ for $\omega_0 \gg k_{\mathrm{B}}T/\hbar$. For broadband driving, the efficiency ratio $\langle \eta \rangle_\sigma / \beta$ depends on how the Drude width $\Gamma$ compares to the Planckian frequency $\omega_{\star}$.
When $\Gamma \ll \omega_{\star}$, the Drude peak is narrow and concentrates spectral weight at low frequencies where $\eta(\omega) \approx \beta$; the efficiency ratio approaches unity.
When $\Gamma \gg \omega_{\star}$, the broad Drude peak spreads spectral weight to frequencies $\omega \gg \omega_{\star}$ where $\eta(\omega) \ll \beta$; the efficiency ratio is parametrically reduced.
The Planckian regime $\Gamma \sim \omega_{\star}$, characteristic of strange metals~\cite{orenstein,Marel2003,basov07,YRS20,Michon2023,DChaudhuri2025,LiangWu}, represents an intermediate case where the relaxation rate is as large as possible while keeping dissipation predominantly in the efficient low-frequency window; see Fig.~\ref{main_fig}.

\textit{Discussion}---
We have established a universal bound relating the Bures distance, and equivalently quantum Fisher information, generated by a time-dependent perturbation to the work dissipated during the same process.
The bound in Eq.~\eqref{eq:main_bound_intro} follows solely from the KMS structure of thermal equilibrium and the positivity of dissipation, with no assumptions about the microscopic Hamiltonian, the presence of quasiparticles, or the functional form of response functions.
The bound reveals that the Planckian timescale $\hbar/k_{\mathrm{B}}T$ governs the efficiency of information encoding --- dissipation at frequencies below this scale is maximally efficient, while high-frequency dissipation is parametrically wasteful. Systems with Planckian relaxation rates represent a critical boundary; they achieve the largest possible dissipation rate while keeping spectral weight in the thermodynamically efficient window. Strange metals thus operate at the edge of optimal information-dissipation efficiency.
This provides a thermodynamic and information-geometric perspective on Planckian phenomenology, complementary to conventional transport analyses.

Our result bears a structural resemblance to TURs~\cite{Barato2015,Horowitz2020}, which bound the precision of current measurements in terms of entropy production.
The classical TUR states that for steady-state currents, $\mathrm{Var}(J) \cdot \Delta S_{\mathrm{irr}} \geq 2 k_{\mathrm{B}} \langle J \rangle^2$.
Our bound $d_B^2 \leq \Delta S_{\mathrm{irr}}/4$ has a similar ``distinguishability $\times$ cost'' structure, motivated by parameter estimation rather than current fluctuations, and to transient linear response rather than nonequilibrium steady states.
Quantum extensions of TURs have been explored~\cite{Hasegawa2020,Guarnieri2019,Carollo2019}, but in a setting that is distinct from the one considered here.

While we focused on current response (relevant for charge transport and optical conductivity), the derivation generalizes straightforwardly to any Hermitian observable $\mathcal{O}$ coupled linearly to an external field. Replacing $J \to \mathcal{O}$ throughout, one obtains
\begin{equation}
[d_B^{(\mathcal{O})}]^2 \leq \frac{\beta}{4} W_{\mathrm{diss}}^{(\mathcal{O})},
\end{equation}
with the efficiency kernel $\eta(\omega)$ unchanged. Each channel has its own dissipation-weighted spectrum $P(\omega)$, and the proximity to saturation characterizes whether dissipation in that channel is Planckian.

There are several future directions that merit further investigation. Perhaps most importantly, our result naturally leads to the fundamental, microscopic question of why Planckian scatterers choose to sit at the edge of optimal efficiency. The behavior near quantum critical points with $\omega/T$ scaling is also likely to reveal interesting insights. Investigating the possible existence of related bounds between dissipation and other information-theoretic quantities, as well as nonlinear response, is an exciting future direction. Finally, an explicit demonstration of near-saturation in response to a calibrated experimental pulse is an exciting direction. 

{\it Acknowledgments---}  DC thanks H. Guo, O. Lesser and D. Mao for useful discussions. DC is supported in part by a grant from the Department of Energy (DE-SC0026112) under the Early Career Research Program. DC acknowledges the use of large language models (ChatGPT 5.2, Gemini 3 and Opus 4.5) for research assistance.

\bibliography{refs}
\end{document}